# An Improved Process for Fabricating High-Mobility Organic Molecular Crystal Field-Effect Transistors

A.P. Micolich [a)], L.L. Bell, and A.R. Hamilton

*School of Physics, University of New South Wales, Sydney NSW 2052, Australia*

In this paper we present an improved process for producing elastomer transistor stamps and high-mobility organic field-effect transistors (FETs) based on semiconducting acene molecular crystals. In particular, we have removed the need to use a silanized Si wafer for curing the stamps and to handle a fragile micron-thickness polydimethylsiloxane (PDMS) insulating film and laminate it, bubble free, against the PDMS transistor stamp. We find that despite the altered design, rougher PDMS surface, and lamination and measurement of the device in air, we still achieve electrical mobilities of order 10 cm$^2$/Vs, comparable to the current state of the art in organic FETs. Our device shows hole conduction with a threshold voltage of order − 9V, which corresponds to a trap density of $1.4 \times 10^{10}$ cm$^{-2}$.

[a)] Corresponding author: mico@phys.unsw.edu.au

# I Introduction

The discovery of inherently conducting polymers by Chiang *et al.* (Ref. 1) sparked intense effort towards the development of 'plastic' electronic devices and integrated circuits for commercial applications.[2] Conducting polymers hold great promise for applications such as flexible displays, organic solar cells, radio frequency ID tags and electronic paper.[3] However, the low electrical mobility of conducting polymers compared to conventional semiconductors (e.g., Si) currently limits their potential for transistor applications.[3,4] For many years, efforts to increase the carrier mobility in organic field-effect transistors (OFETs) have focused on improving the structural ordering at the molecular level, chiefly through the molecular design of polymers (e.g., regioregular poly(3-hexylthiophene)) that self-organize into well-ordered structures.[5] This approach has led to conducting polymer films with electrical mobility as high as 0.1 cm$^2$/Vs – only an order of magnitude smaller than typically found in amorphous Si.[6]

While this approach provides increased mobility through short-range structural ordering, the chain length of conducting polymers limits their long-range order, and hence their mobility. Higher mobilities can be achieved using shorter conjugated oligomers, such as the acenes (e.g., pentacene, tetracene, etc.), which are well known to form highly ordered molecular crystals.[4] Acene-based transistors have typically been produced by thermally evaporating an oligoacene layer onto a prefabricated Si FET structure consisting of a doped Si substrate, which acts as the gate, a thermally-grown SiO$_2$ layer as the insulator, and evaporated metal (e.g., Ti/Au or Ag) source/drain contacts.[7] However, oligoacene films deposited onto such a structure are polycrystalline because growth spreads from various nucleation sites on the substrate surface, limiting the mobility of the deposited films. The contacts, whether deposited before or after the oligoacene film, lead to additional defects/damage, further limiting the mobility. Hence the present highest mobility for Acene-on-Si OFETs is of order 1 cm$^2$/Vs.[7]

An alternative approach, leading to even higher mobilities, is to produce OFETs based on high-quality, freestanding organic molecular crystals.[8] Single organic molecular crystals of oligoacenes with mm-dimensions are easily grown using the physical vapour growth technique pioneered by Laudise *et al.* (Ref. 9). Podzorov *et al.* (Ref. 10) recently demonstrated an organic FET based on a freestanding rubrene crystal using Ag contacts and gate and a parylene insulator deposited by low-temperature CVD to obtain mobilities up to ~10 cm$^2$/Vs at room temperature. A major advance was made by Sundar *et al.*, who reported a room temperature mobility of 15 cm$^2$/Vs from OFETs produced by laminating a rubrene crystal against a polydimethylsiloxane (PDMS) elastomer stamp containing the gate, insulator and source and drain contacts.[11] This approach solved some key problems of previous devices by eliminating the multiple processing steps that contaminate and strain the fragile crystal, thereby keeping the interface between the crystal and dielectric as clean as possible. Recent research has also shown that laminated contacts to organic semiconductors provide contact resistances an order of magnitude lower than obtained by evaporating metal directly onto an organic semiconductor.[12]

The original method for making elastomeric transistor stamps[11] is somewhat complex and involves a number of steps that can limit the yield of working devices. In this paper, we discuss an alternative fabrication strategy for producing elastomer transistor stamps that requires fewer resources and processing steps without compromising the transport properties of the resulting OFETs. We demonstrate this by producing rubrene OFETs with room temperature mobilities of order 10 cm$^2$/Vs, which are comparable to the best results achieved to date.[11,13]

The paper is arranged as follows: In Section II-A we briefly summarise the existing technique used by Sundar *et al.* to produce elastomer stamps. We then introduce our alternate technique, compare it with Ref. 11, and discuss some of the issues related to reliable device fabrication in Section II-B. In Section II-C we present the electronic properties of OFETs produced using our technique. Finally, in Section III we review the key findings of our work.

## II  Elastomer Stamp OFET Fabrication

In this section we will briefly review the method reported by Sundar *et al.* (Ref. 11) for making OFETs with an elastomer stamp, prior to introducing our alternate technique.

### II-A  Elastomer stamp method used by Sundar *et al.*

The elastomer stamps used by Sundar *et al.* are produced by curing a thin layer of Sylgard 184 PDMS (Dow Corning) against a cleaned Si wafer coated with a monolayer of $CF_3(CF_2)_6(CH_2)_2SiCl_3$ (Gelest Inc.). A 2 nm Ti / 30 nm Au / 3 nm Ti gate layer is then deposited on the cured layer by thermal evaporation. A separate thin PDMS layer is prepared by spin-coating the silanized Si wafer with PDMS thinned with trichloroethylene. This second cured layer (~3 μm thick) is laminated over the gate layer to form the gate insulator, and 2 nm Ti / 20 nm Au source/drain contacts are then deposited on it by thermal evaporation. The final step is the lamination of a rubrene crystal against the exposed source and drain contacts, forming a structure similar to that shown in Fig. 1(a).

### II-B  Our Alternate Fabrication Strategy

Our transistor stamp consists of a PDMS substrate onto which we deposit a Ti/Au gate layer, a thin PDMS gate insulator, and finally a set of Ti/Au source and drain contacts. The resulting transistor stamp is laminated against a rubrene crystal to form the OFET device shown schematically in Fig. 1(a). Our devices differ from those in Ref. 11 as discussed below:

1) **Preparation of the Substrate:** The casting of PDMS layers against Si wafers provides the very flat, even surfaces important to achieving high resolution soft lithography and layer bonding in microfluidics. However, such flatness is not essential in elastomer transistor stamps because the stamp surface deforms to achieve Van der Waals contact with the crystal surface, which is significantly rougher than the Si wafer surface. Hence, we decided to eliminate the use of a silanized Si-wafer and seek a cheaper and easier material for the casting process. Our use of teflon (polytetrafluoroethylene) was motivated by the use of a heavily-fluorinated silane monolayer to avoid the PDMS from sticking to the Si in Ref. 11. We prepared casting 'chucks' by machining flat ends (high rotation/low travel speed to minimise surface roughness) onto 20 mm lengths of 1" diameter teflon rod. We deposit PDMS onto the chuck and allow it to cure at room temperature for 24 hours. This produces a relatively thick (~ 3 mm) substrate, as shown in Fig. 1. The surface cured against the PTFE chuck is too rough for use in devices, however, the upper surface of the cured PDMS substrate is very smooth and is suitable for device fabrication Although the PDMS cure time can be reduced from ~24 hours at room temperature to < 1 hour at temperatures $T > 100°C$, we cure at room temperature to provide the smoothest possible PDMS surface for the subsequent device fabrication stages.

2) **Evaporation of the Ti/Au gate layer:** Sundar *et al.* deposit Ti/Au gate and contact layers after the PDMS is exposed to a low power (~10W) $O_2$ plasma for several seconds. The aim of the Ti layers and $O_2$ plasma treatments are to improve the adhesion between the Au and PDMS.[11] We found that the second Ti overlayer and the $O_2$ plasma treatments, although desirable, are not essential to the operation of our OFETs.[14] Instead, there are two more important parameters affecting the electrical continuity, resistance and robustness of metal layers deposited on PDMS: The first is the thickness of Au used. We find that thicker layers (3 nm Ti / 40 nm Au) are needed with our stamps, possibly due to the higher surface roughness of our air-cured PDMS surface compared to a PDMS surface cured against a Si wafer.

The second and more important issue is the PDMS surface temperature during metal deposition. If care is not taken to minimise heating of the PDMS surface during deposition, then the resulting film develops micron-scale ripples due to the difference in thermal expansion coefficient between PDMS and Ti/Au.[15] This leads to gates/contacts with a frosty/opalescent appearance and either no electrical continuity or a very high resistance (> 10 MΩ), making them unsuitable for device applications. There are three measures that we have found useful in minimising the heat load during evaporation. The first is to maximise the source-sample separation, however eventually increased separation becomes detrimental because it increases the evaporation time and thus the radiant heat-load on the sample, we presently use a separation of 120 mm. The second is to mount the sample on a water-cooled stage. The final measure is to evaporate the Ti as slowly as possible (rate ~ 0.1 nm/s) but to evaporate the Au as quickly as possible (> 1.2 nm/s). When combined these three measures allow us to obtain very high-quality Ti/Au films (see Fig. 1(b)) with low resistance (10-100 Ω) and a typical yield exceeding 90%.

3) **The PDMS gate insulator:** PDMS layers of order microns thick are difficult to handle and laminate over the gate without trapping small bubbles of air between the two PDMS surfaces. This approach also requires a number of additional processing steps. To eliminate these, we spin-coat the PDMS gate insulator directly onto the PDMS stamp after gate deposition. We increased the thickness of the insulator from ~3 μm to ~ 10 μm to account for the slightly rougher surface in our PDMS substrates. This removes the need to thin the PDMS with a solvent, but results in higher FET operating voltages. The unthinned PDMS is quite viscous and requires a two-stage spinning process to avoid streaking: we use 500 rpm for 30 sec followed by a higher speed spin (see below) for 1 min, with an overnight cure at room temperature. When spin-coating the PDMS insulator over the metal gate layer, it is essential to leave a small part of the gate layer uncoated to allow electrical contact to the gate. To achieve this, we 'mask' the corner of the gate layer by covering it with a small square of thin cured PDMS film, which is removed immediately after the PDMS insulator layer is spun-on.

The thickness of the PDMS insulator is controlled by the spin-speed, with speeds between 3000 and 9000 rpm yielding film thicknesses between 5 and 25 μm. The measurement of the thickness of spun-on PDMS layers is not straightforward. Due to the elastomeric nature of PDMS, mechanical probe techniques give unreliable results. We have found two methods that do give reliable thickness results: The first is a capacitive technique and the second involves optical interference measurements. For the capacitive technique we create a structure that consists of a PDMS substrate, a Ti/Au gate layer, a spun-on PDMS insulator and another Ti/Au gate layer, this time deposited perpendicular to the first gate layer. The capacitance of the PDMS gate insulator layer can then be measured at a frequency of 100 Hz using a lock-in amplifier,[16] and its thickness determined knowing the overlap area of the two metal films and the dielectric constant of PDMS $\varepsilon = 2.66$. An advantage of this technique is that one can measure the capacitance per area for a nominally identical PDMS film, allowing more accurate estimates of capacitances in the FETs. Optical measurements were performed (to confirm the capacitive results) on structures consisting of substrate, gate layer and PDMS film. The thickness is measured based on the thin-film interference of the PDMS film using a laser tunable between $\lambda = 600 - 800$ nm. The measured thickness vs spin-speed for PDMS films spun on PDMS substrates is shown in Fig. 2, with capacitive measurements in red and optical measurements in blue, giving good agreement.

4) **Other notable aspects of device fabrication:** The source and drain contacts (3 nm Ti / 40 nm Au) are deposited through an aligned shadow mask by thermal evaporation. We grow our rubrene single-crystals using a physical vapor transport method,[9] with a 30 mL/min flow of high-purity Ar and a source temperature of 300°C. The crystals deposit below 220°C forming hexagonal flakes with dimensions up to ~ 8 mm across and of order 0.1 mm thick; with a typical growth time of 24 hours we get 2-3 useful crystals (at least a few mm square ) per growth run. The completed devices are produced by laminating a fresh rubrene crystal against the stamp.

**II-C   Electronic Properties**

The completed devices are connected to the external measurement circuit using probe needles, as shown in Fig. 1(c). The device was measured using a Keithley 6517A electrometer to apply the source-drain bias $V_{ds}$ and measure the resulting source-drain current $I_{ds}$, and a Keithley 2410 source-measure unit to apply the gate bias $V_g$ and continuously monitor the gate leakage current $I_g$. Measurements were performed in air at room temperature inside a light-tight grounded metal box. The OFETs we have measured have a large active region, with a channel length $L$ = 2.6 mm and width $W$ = 1.5 mm. The gate insulator is 10 μm thick, with a measured capacitance $C_g$ ~ 0.25 nF/cm$^2$. The expected breakdown voltage for this insulator is 212 V, and over the ±50 V range of our measurements (maximum hole density of 8 × 10$^{10}$ cm$^{-2}$ corresponding to 4 × 10$^{-4}$ holes/molecule), the gate leakage current did not exceed 10 nA.

In Fig 3(a)/(b) we present the source-drain characteristics and transfer characteristics (log-scale) for one of our rubrene/PDMS OFETs. The devices show typical p-type behaviour (conduction via holes), consistent with previous studies of single-crystal rubrene FETs,[8,10,11] and many other organic FETs[4], with source-drain currents as high as several μA easily obtained. We do not observe ambipolar conduction in these devices. From the data in Fig. 3(b), we obtain subthreshold slopes ranging from 4 to 8 V/decade, which after correcting for the dielectric capacitance $C_g$, gives a maximum normalised sub-threshold slope of 1.6 V.nF/decade.cm$^2$, considerably better than found in thin-film pentacene OFETs, and comparable to that found in other rubrene molecular crystal OFETs.[8,10] The on-off ratio for our devices exceeds 10$^4$.

Figure 4 shows the low-$V_{ds}$ transfer characteristics (linear scale) from which we obtain the device threshold voltage. By extrapolating the linear regions of the low-$V_{ds}$ transfer characteristics for different $V_g$ back to where they converge at $I_{ds} = 0$, as shown in Fig. 4, we obtain a threshold voltage $V_T = -9.1 \pm 0.5$ V, which corresponds to a trap density of $p = 1.4 \times 10^{10}$ cm$^{-2}$. Such negative threshold voltage behaviour is commonly[8,13,17-19] (but not always[10,11]) observed in molecular crystal OFETs, and is typically < 15 V, as found in our devices. Surprisingly, despite the rougher surfaces of our stamps, the solid dielectric to solid crystal interface, large channel dimensions and the assembly and measurement of our devices under ambient conditions, we find typical mobilities on the range 5-10 cm$^2$/Vs for our devices. Figure 5 shows a plot of the mobility (left axis) and calculated hole density (right axis) as a function of gate bias $V_g$. For high gate biases ($V_g < -22$ V) we find a density-independent mobility, consistent with behaviour commonly observed in single-crystal organic FETs.[8]

## III Discussion and Conclusions

In this paper, we have presented an alternate way of making PDMS transistor stamps. Our aim was to simplify the fabrication process without compromising the quality of resulting devices. In particular, we have removed the need to prepare and use a silanized Si wafer for curing the stamps, and to handle a fragile micron thickness PDMS film and laminate it, bubble free against the PDMS stamp. We achieve this by using the top surface of a layer cured on a flat teflon surface, and a spun-on PDMS insulator layer. We have also found that water-cooling and controlled evaporation is key to achieving flat low-resistance metal layers. We find that despite the simpler design, rougher PDMS surface and the lamination and measurement in air, we still achieve mobilities of order 10 cm$^2$/Vs. Our device shows hole conduction with a threshold voltage of − 9.1 V, corresponding to a trap density of $1.4 \times 10^{10}$ cm$^{-2}$.

**Acknowledgements:** We thank Suhra Ilyas and Michael Gal for help with the optical measurement of PDMS insulator thickness, and Jack Cochrane and Ali Rashid for construction of the crystal growth furnaces. This work was funded by the Australian Research Council (ARC) under DP0346279, LE0239044, and UNSW internal grants. APM acknowledges the award of an ARC Australian Postdoctoral Fellowship.

16. The voltage reference is fed onto one plate of the capacitor and the other plate is connected to the current input (virtual ground) allowing the capacitive reactance (90° component) and the insulator resistance (0° component) to be measured simultaneously.

17. V. Podzorov, V.M. Pudalov and M.E. Gershenson, *Appl. Phys. Lett.* **82**, 1739 (2003).

18. V. Podzorov, E. Menard, A. Borissov, V. Kiryukhin, J.A. Rogers and M.E. Gershenson, *Phys. Rev. Lett.* **93**, 086602 (2004).

19. C.R. Newman, R.J. Chesterfield, J.A. Merlo and C.D. Frisbie, *Appl. Phys. Lett.* **85**, 422 (2004).

**Figure Captions**

Figure 1: (a) Schematic cross-section of a rubrene/PDMS OFET showing the transistor stamp with gate, PDMS insulator, source and drain contacts and a rubrene crystal, (b) Photograph of a Ti/Au gate layer deposited on PDMS with the surface heat-load minimised, and (c) Photograph of a completed device during measurement.

Figure 2: PDMS insulator film thickness, as deposited on a PDMS substrate, as a function of spin speed. Data obtained using a capacitive method (red) and an optical interference method (blue).

Figure 3: (a) Source-drain characteristics $I_{sd}$ vs $V_{sd}$ at various $V_g$ ranging from –50 V (top) to +20 V (bottom) in steps of 5 V, and (b) transfer characteristics $I_{sd}$ (log-scale) vs $V_g$ at various $V_{sd}$ ranging from +45 V (top) to +5 V (bottom) in steps of 5 V for our rubrene/PDMS OFET. Our FETs show typical p-type FET behaviour.

Figure 4: Small source-drain bias transfer characteristics $I_{sd}$ (linear scale) vs $V_g$ at small $V_{sd}$ ranging from +1.1 V (top) to +0.1 V (bottom) in steps of 0.1 V. The red dashed lines in (b) are guides to the eye regarding the determination of the threshold voltage $V_T = -9.1 \pm 0.5$ V.

Figure 5: Electrical mobility (left axis – red diamonds) and hole density (right axis – blue line) as a function of gate bias $V_g$ showing density-independent mobility for $V_g < -22$ V. The black lines are guides to the eye.

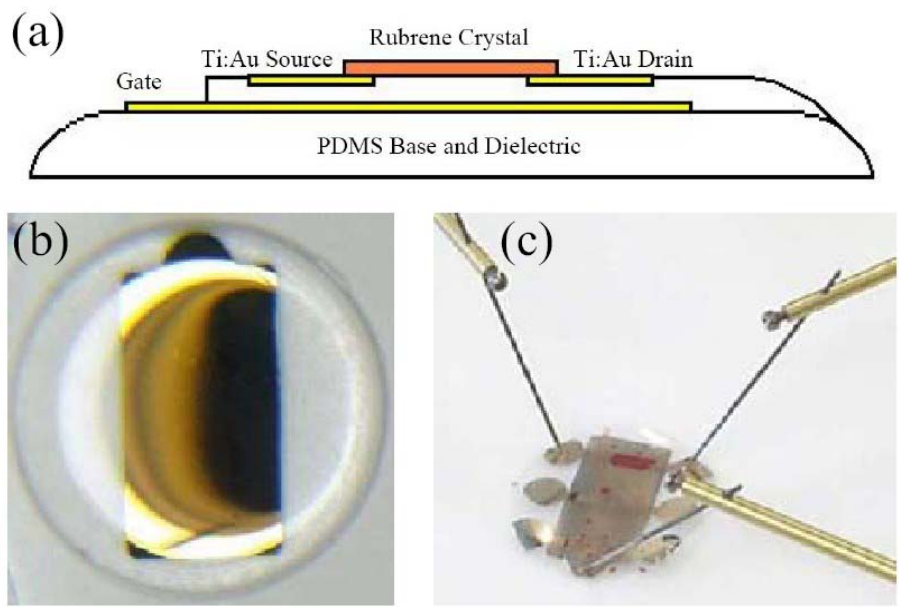

Figure 1 of A.P. Micolich *et al.*, An Improved Process for Fabricating High-Mobility Organic Molecular Crystal Field-Effect Transistors. (Color Online)

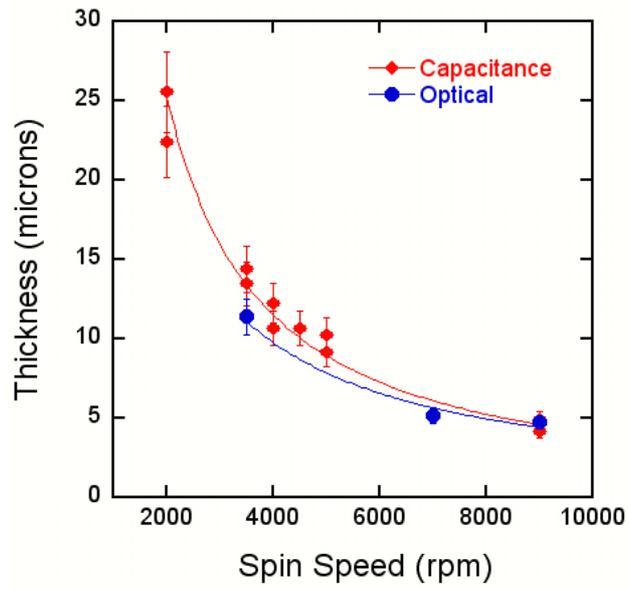

Figure 2 of A.P. Micolich *et al.*, An Improved Process for Fabricating High-Mobility Organic Molecular Crystal Field-Effect Transistors. (Color Online)

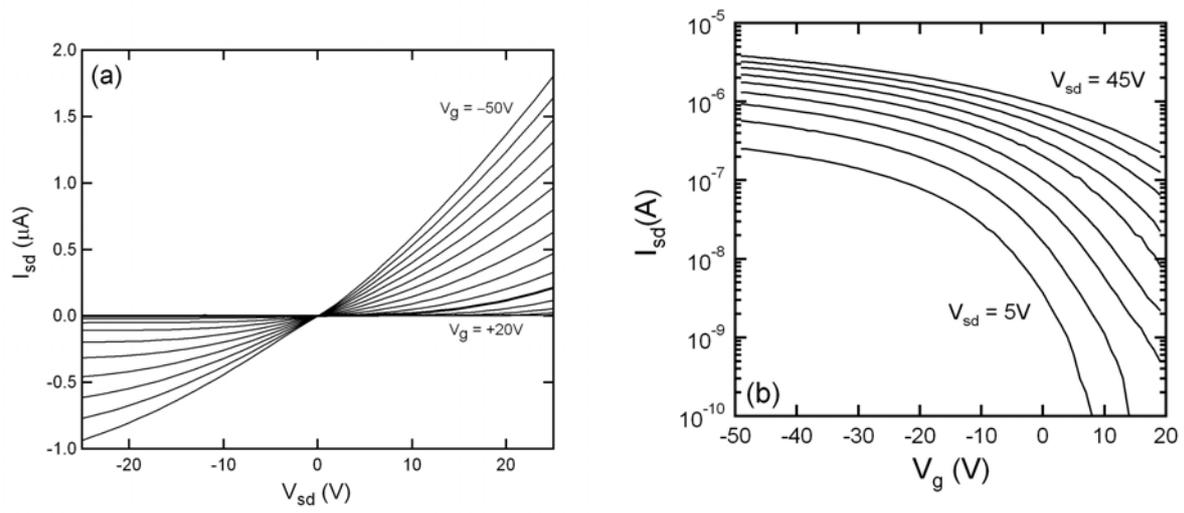

Figure 3 of A.P. Micolich *et al.,* An Improved Process for Fabricating High-Mobility Organic Molecular Crystal Field-Effect Transistors. (Color Online)

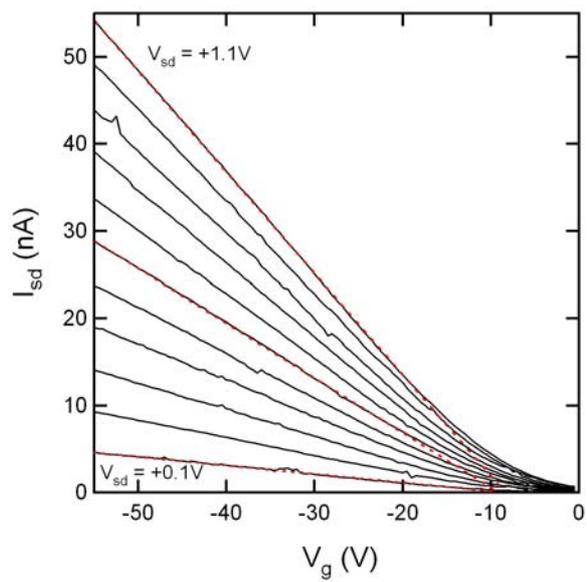

Figure 4 of A.P. Micolich *et al.*, An Improved Process for Fabricating High-Mobility Organic Molecular Crystal Field-Effect Transistors.

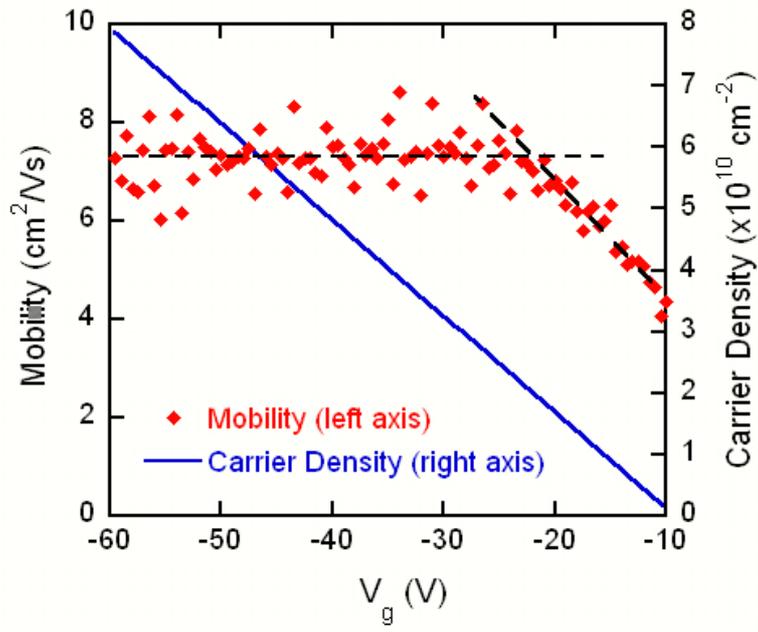

Figure 5 of A.P. Micolich *et al.,* An Improved Process for Fabricating High-Mobility Organic Molecular Crystal Field-Effect Transistors. (Color Online)